\documentclass[mathleft
]{an}
\usepackage{graphicx}
\usepackage{times}
\usepackage{amssymb}
\begin{document}

\Pagespan{1}{}
\Yearpublication{2012}%
\Yearsubmission{2012}%
\Month{}%
\Volume{}%
\Issue{}%

\title{Milky Way simulations: the Galaxy, its stellar halo and its satellites
  -- \\ insights from a hybrid cosmological approach}

\author{Gabriella De Lucia\thanks{\email{delucia@oats.inaf.it}\newline}
}
\titlerunning{Milky Way simulations}
\authorrunning{G. De Lucia}
\institute{INAF - Astronomical Observatory of Trieste, via G.B. Tiepolo 11, 
        I-34143 Trieste, Italy}

\received{2012}
\accepted{2012}
\publonline{later}


\abstract{Our `home galaxy' - the Milky Way - is a fairly large spiral galaxy,
  prototype of the most common morphological class in the local Universe.
  Although being only {\it a galaxy}, it is the only one that can be studied in
  unique detail: for the Milky Way and for a number of members of the Local
  Group, a wealth of observational data is available about the ages and
  chemical abundances of their stars. Much more information is expected to come
  in the next few years, from ongoing and planned spectroscopic and astrometric
  surveys, providing a unique benchmark for modern theories of galaxy
  formation. In this review, I will summarize recent results on the formation
  of our Milky Way, its stellar halo, and its satellite galaxies. I will focus,
  in particular, on results obtained in the framework of hybrid models of
  galaxy formation, and refer to other reviews in this issue for studies based
  on hydrodynamical simulations.}  \maketitle

\section{Introduction}
\label{sec:intro}
Our Galaxy - the Milky Way - is a fairly large spiral galaxy consisting of four
stellar components: most of the stars are distributed in a thin disk, exhibit a
wide range of ages, and are on high angular momentum orbits. About 10-20 per
cent of the mass in the thin disk resides in a distinct component which is
referred to as the thick disk. This is composed of old stars, that have on
average lower metallicity than those of similar age in the thin disk, and are
on orbits of lower angular momentum. The Galactic bulge is dominated by an old
and relatively metal-rich stellar population, with a tail to low abundances.
Finally, a tiny fraction of the total stellar mass (only a few percent) resides
in the stellar halo, which is dominated by old and metal poor stars on low
angular momentum orbits.

Although the Milky Way is only {\it one galaxy}, it is our {\it home galaxy}
and, as such, it is the only one that we can study in unique detail. Over the
past decades, accurate measurements of physical properties (ages and
metallicities) \linebreak and kinematics have been collected for a significant
number of individual stars. Many more data are coming in the next future, from
ongoing or planned astrometric and spectroscopic surveys. This wealth of
observational data and details provide a unique benchmark for modern theories
of galaxy formation and evolution.

Historically, chemical and kinematic information for \linebreak stars and
stellar systems in the solar neighbourhood were used as a basis to formulate the
first galaxy formation models. Eggen, Lynden-Bell \& Sandage (1962) analysed a
sample of $\sim 200$ dwarfs and showed that stars with the lowest metallicity
tended to move on highly elliptical orbits. The data were interpreted as
evidence that the oldest stars in the galaxy were formed out of gas collapsing
from the halo onto the plane of the galaxy, on relatively short time-scales (a
few times $10^8$~years). About one decade later, Searle \& Zinn (1978) analysed
a sample of $\sim 200$ giants and $\sim 20$ globular clusters finding no radial
abundance gradients. These observations led to the formulation of a different
scenario in which the stellar halo forms through the agglomeration of many
sub-galactic fragments, that may be similar to the surviving dwarf spheroidal
satellites (dSphs) of our Galaxy.

The Searle \& Zinn scenario appears to be in qualitative agreement with
expectations from the hierarchical cold dark matter model (CDM), and evidence
in support of the proposed picture has mounted significantly over the past
years. This ranges from the detection of significant clumpiness in the phase
space distribution of halo and disk stars, to the detection of satellite
galaxies caught in the act of tidal disruption. Some problems, however,
remain. A first one was pointed out by Shetrone, Cote \& Sargent (2001) who
noted that stars in the Local Group dSphs have lower alpha abundances than
stars in the stellar halo. These observations suggest that the Galactic stellar
halo cannot result from the disruption of satellite galaxies similar to the
{\it surviving} dSphs of the Local Group. The difficulty can be overcome if
surviving satellites are intrinsically different from satellites that have
contributed to the formation of the stellar halo (I will discuss this in more
detail in Section 4).

A potentially more serious problem with the Searle \& Zinn scenario was pointed
out by Helmi et al. (2006) who found a lack of stars of low metallicity in four
nearby dSphs (Sculptor, Sextans, Fornax, and Carina), suggesting that the
material that gave origin to these galaxies was enriched prior to their
formation. In addition, Helmi et al. found that the metal-poor tail of the dSph
metallicity distribution is significantly different from that of the Galactic
halo, arguing that the progenitors of present day dSphs are fundamentally
different from the building blocks of our Galaxy, even at earliest epochs. The
picture has, however, changed recently as many studies have detected very
metal-poor stars both in classical and in ultra-faint dSphs (e.g. Kirby et al
2008; Frebel et al. 2010).

Another element of crisis with respect to the current standard cosmological
paradigm is provided by the so called {\it missing satellite problem}, i.e. the
observation that substructures resolved in a galaxy-size DM halo significantly
outnumber the satellites observed around the Milky Way \linebreak (Klypin et
al. 1999; Moore et al. 1999). It has long been realized, however, that this
problem might have an astrophysical solution. In particular, the presence of a
strong photoionizing background, possibly associated with the reionization of
the Universe, can suppress accretion and cooling in low-mass haloes thereby
suppressing the formation of small galaxies (e.g. Efstathiou 1992, Okamoto, Gao
\& Theuns 2008 and references therein). Galaxy formation in small haloes can
also be significantly suppressed by supernovae driven winds, and the
combination of these two physical processes can bring the predicted number of
luminous satellites in agreement with observational data (e.g. Kauffmann, White
\& Guiderdoni 1993; Bullock, Kravtsov \& Weinberg 2000; Benson et al. 2002).

Another potential problem with the satellite galaxies has emerged in more
recent years: the dwarf satellite galaxies of the Milky Way span several orders
of magnitude in luminosity. Yet, the mass enclosed within a small radius (300
or 600 pc) appears to be roughly constant (Strigari et al. 2008), suggesting
the existence of a minumum mass scale which is not present in the primordial
CDM power spectrum of density perturbation. Therefore, in the framework of the
current favourite cosmological model, such a minumum scale can only result from
astrophysical processes. 

The discovery of a new population of ultra-faint satellites in the past few
years has led to a renewed interest in the physics of dwarf galaxy
formation. It should be noted, however, that this discovery did not alleviate
the original missing satellite problem, as all the newly discovered satellites
are fainter than the classical ones. New impetus to the field has also been
given by the completion of extremely high resolution $N$-body simulations of
Milky-Way size haloes (Springel et al. 2008, Diemand et al. 2008).

Space limit does not allow an exhaustive review of all studies related to the
formation and evolution of the Milky Way and its satellite population. In the
following, I will therefore focus on the most recent results obtained in the
framework of {\it hybrid models of galaxy formation} (see next section). I
refer to the reviews by C. Scannapieco and V. Springel for related discussions
focused on hydrodynamical simulations.

\section{Simulations and Galaxy Formation Models}

The formation and the evolution of the baryonic component of galaxies is
regulated by a number of non-linear processes operating on vastly different
scales (e.g. shocking and cooling of gas, star formation, feedback by
supernovae and active galactic nuclei, etc.). Most of these processes are
poorly understood, even when viewed in isolation. The difficulties grow
significantly when considering that, in the real Universe, these processes act
together in a complex network of actions, back-reactions, and self-regulation.

In recent years, different approaches have been used to link the observed
properties of luminous galaxies to the dark matter haloes within which they
reside. Among these, semi-analytic models have developed into a powerful and
widely used tool to study galaxy formation in the framework of the currently
standard model for structure formation. In this approach, the evolution of the
baryonic component is modelled invoking simple, yet physically and
observationally motivated {\it prescriptions}.  These techniques find their
seeds in the pioneering work by White \& Rees (1978), but have been
substantially extended and refined in the last years by a number of different
groups (recent work includes De Lucia \& Blaizot 2007, Monaco et al. 2007,
Somverville et al. 2008, Benson \& Bower 2010, Guo et al. 2011).

Modern semi-analytic models of galaxy formation take advantage of high
resolution N-body simulations to specify the location and evolution of dark
matter haloes - which are assumed to be the birthplaces of luminous galaxies.
Using this {\it hybrid} approach, it is possible not only to predict observable
physical properties such as luminosities, metallicities, star formation rates,
etc., but also to provide fully dynamically consistent spatial and kinematical
information for all model galaxies. This allows more accurate and
straightforward comparisons with observational data to be carried out.

Most of the results discussed below are based on $N$-body re-simulations of
`Milky-Way' haloes: first, a cosmological simulation of a large region is used
to select a suitable target halo.  The particles in this halo and its
surroundings are then traced back to their initial Lagrangian region and
replaced with a larger number of lower mass particles. Outside the
high-resolution region, particles of mass that increases with distance are
used, so that the computational effort is concentrated on the region of
interest while a faithful representation of the large scale density field is
maintained.

\begin{figure}
\begin{center}
\includegraphics[width=79mm,height=79mm]{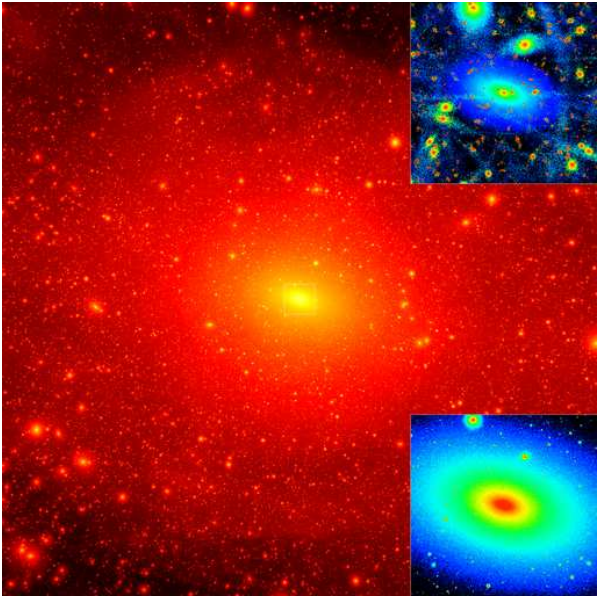}
\caption{From Diemand et al. (2008): projected dark matter density-square map
  of the Via Lactea II simulated halo. The region shown corresponds to a 800
  kpc cube. The insets focus on an inner 40 kpc cube, in local density
  (bottom), and in local phase space density (top).}
\label{fig:lacteaII}
\end{center}
\end{figure}

In the following, I will present results obtained mainly from two sets of
re-simulations. For the GA series (Stoehr et al. 2003), a Milky-Way halo was
selected as a relatively isolated halo which suffered its last major merger at
$z>2$, and with approximately the correct peak rotation velocity. The selected
halo was then re-simulated at a series of four progressively higher resolution,
with the highest resolution simulation (the GA3) using a particle mass $m_{\rm
  p} = 2.947\times10^5\,{\rm M}_{\sun}$ and a gravitational softening $\epsilon
= 0.18\,h^{-1}$kpc. In the Aquarius project, carried out by the Virgo
Consortium, six galactic dark matter haloes with masses comparable to that of
the Milky Way were extracted from a cosmological box and re-simulated at
varying levels of resolution (Springel et al. 2008). In the following, I will
present results obtained by coupling the second highest level of resolution of
the Aquarius haloes ($m_{\rm p}$ ranges from $6.447\times 10^3$ to $1.399\times
10^4\,{\rm M}_{\sun}$, and $\epsilon = 65.8$~pc) with independently developed
semi-analytic models. Finally, the Via Lactea II simulation has been recently
completed by Diemand et al. (2008). A projected dark matter density map from
this simulation is shown in Fig.~\ref{fig:lacteaII} (this is one galaxy-mass
halo, simulated using particles of mass $m_{\rm p}=4.1\times10^3\,{\rm
  M}_{\sun}$, and $\epsilon=40$~pc).

\section{The Galaxy}

The formation and evolution of different stellar components of the Milky Way
has been studied in detail over the past decades, using a combination of
$N$-body simulations and analytic models. In the framework of cosmological
hydrodynamical simulations, the formation of realistic disk galaxies has long
been considered, and still largely remains, a challenge. This is due to a
combination of numerical effects and limitations of the treatment of baryonic
processes. Much progress has been made, highlighting the crucial role played by
feedback in preventing overcooling and regulating the assembly of galaxies in
order to avoid catastrophic angular momentum losses (see e.g. Scannapieco et
al. 2011 and references therein).

\begin{figure}
\includegraphics[width=80mm,height=67mm]{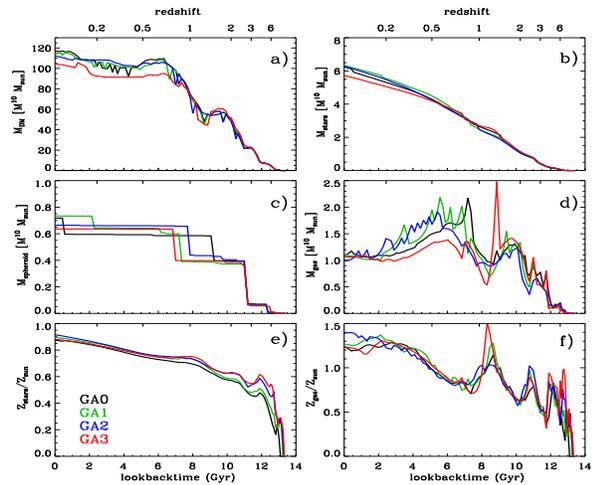}
\caption{From De Lucia \& Helmi (2008): evolution of the dark matter mass
  (panel a), total stellar mass (panel b), spheroid mass (panel c), cold gas
  mass (panel d), stellar metallicity (panel e), and gas metallicity (panel f)
  for the model Milky Way galaxies in the four simulations of the GA series
  (lines of different colours).}
\label{fig:mgrowth}
\end{figure}

In the framework of hybrid models of galaxy formation, a recent work has been
carried out by De Lucia \& Helmi (2008) by taking advantage of the GA series.
Figure~\ref{fig:mgrowth} is reproduced from this work, and shows the evolution
of different mass and metallicity components for the model Milky Way galaxies
in all four simulations of the series (black corresponds to the lowest
resolution simulation, and red to the highest resolution one). The histories
are constructed by tracking the evolution of the main progenitor, obtained by
linking the galaxy at each time-step to the progenitor with the largest stellar
mass. For this model Milky Way, the galaxies merging onto the main branch have
stellar masses that are much smaller than the current mass of the main
progenitor, over most of the galaxy's life-time, explaining the smooth increase
of the stellar mass component (panel b). The mass of the spheroidal (panel c)
component grows in discrete steps as a consequence of the assumption that it
grows during mergers and disk instability events (see original paper for
details). In this particular model, disk instability contributes about half of
the final stellar mass in the spheroidal component.  Minor mergers contribute
to the other half, and all occur at later times with respect to the disk
instability episodes.

Interestingly, the model provides consistent evolutions for all four
simulations, despite a very large increase in numerical resolution (a factor
$\sim 800$ between GA0 and GA3). Some panels (e.g. panels b, e, and f) do not
show convergence of the results. This is driven by the lack of complete
convergence in the $N$-body simulations (panel a shows a clear difference
between GA3 and the lower resolution simulations).  The reference model used
in the work by De Lucia \& Helmi is in relatively good agreement with
observational measurements (but the mass of the spheroid is slightly lower and
the gas content higher than the observational estimates).

\begin{figure}
\includegraphics[width=83mm,height=67mm]{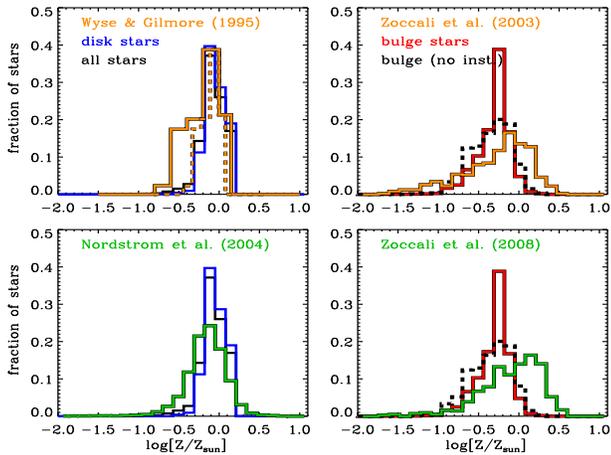}
\caption{From De Lucia \& Helmi (2008): metallicity distribution for stars in
  the disk (blue histograms, left panels) and spheroid (red histograms, right
  panels) of the model Milky Way from the simulation GA3. The solid black
  histograms in the left panels show the metallicity distribution for all stars
  in the model galaxy, while the dashed black histograms in the right panels
  show the metallicity distribution of stars in the spheroidal component in a
  model with disk instability switched off. Orange and green histograms show
  different observational measurements.}
\label{fig:metdistr}
\end{figure}

The metallicity distributions of the stars in the disk and spheroid of the
model Milky Way from the simulation GA3 are shown in
Fig.~\ref{fig:metdistr}. The left panels show the metallicity distribution of
all stars (black histograms) and of the stars in the disk (blue histograms),
compared to different observational measurements. The figure shows that the
metallicity distribution of disk stars in the model Milky Way peaks at
approximately the same value as observed, but exhibits a deficiency of low
metallicity stars. In comparing the model and observed metallicity
distributions, two factors should be taken into account: (1) the observational
metallicity measurements have some uncertainties ($\sim 0.2$~dex) that broaden
the true underlying distribution, and are not included in the theoretical
prediction; (2) the metallicity estimate used in the observational studies is
an indicator of the {\it iron} abundance, which is mainly produced by
supernovae Ia and not well described by the instantaneous recycling
approximation adopted in the model by De Lucia \& Helmi. In order to estimate
the importance of this second caveat, they have converted the measured [Fe/H]
into [O/H] using a linear relation, obtained by fitting observational data for
thin disk stars. The result of this conversion is shown by the dashed orange
histogram in the top panel of Fig.~\ref{fig:metdistr}, and is much closer to
the modelled log[Z/Z$_\odot$] distribution.

The right panels in Fig.~\ref{fig:metdistr} show the metallicity distribution
of the spheroid stars in the fiducial model (red histogram) and in a model
where the disk instability channel is switched off (dashed black histograms).
The metallicity distribution of the model spheroid peaks at lower value than
observed, and it contains a fraction of high metallicity stars that is smaller
than that observed.  Comparison between the solid red histogram and the black
dashed histogram shows that disk instability is responsible for the pronounced
peak around Log[Z/Z$_{\sun}$]$\sim -0.25$, due to the transfer of a large
fraction of disk stars into the spheroidal component. The same caveat discussed
above applies to the metallicity distribution of the spheroidal
component. Note, however, that a conversion from [Fe/H] to [O/H] of the
observed metallicity scale would bring most of the observed bulge stars to
[O/H]$\gtrsim -0.2$, suggesting that the model spheroid in the work by De Lucia
\& Helmi is significantly less metal rich than the observed Galactic bulge.

The use of the instantaneous recycling approximation represents clearly one
important limitation of the study discussed above. Work is ongoing to relax
this approximation and include a more detailed treatment of the chemical
enrichment, by taking into account the lifetime of stars of different
mass. Detailed chemical models for the Milky Way are available (see
e.g. Matteucci 2008 and references therein). Unfortunately, however, these
analytic and numerical models are often carried out under the assumption that
the galaxy evolved in isolation (the models are not embedded in a cosmological
framework), and often assuming ad hoc initial conditions. 
\section{The stellar halo}

As mentioned above, the stellar halo is the Milky Way stellar component that
contains the smallest fraction of its stars. Yet, it is arguably the component
that carries the most useful information about the evolutionary history of the
Galaxy because it contains its most metal-poor stars. The origin and structure
of the stellar halo has been studied by several authors, using different
techniques (for a review, see Helmi 2008). These include cosmological
simulations with and without baryonic physics, and phenomenological models,
usually in combination with $N$-body simulations that provide the dynamical
history of the system.

For example, Bullock \& Johnston (2005) combined \linebreak mass accretion
histories of galaxy-size haloes, constructed using the extended Press-Schechter
formalism, with chemical evolution models for individual satellites. For each
accretion event, they run $N$-body simulations (dark matter only), following
the dynamical evolution of the accreted satellites, which are placed on orbits
consistent with those found in cosmological simulations. A variable
mass-to-light ratio was assigned to each dark matter particle, and chemical
evolution was modelled by considering both Type II and Type Ia supernovae. The
build up and chemical properties of the stellar halo in these models was
analysed by Font et al. (2006). These authors showed that the model reproduces
the systematic differences between the chemical abundances of stars in
satellite galaxies and those in the Milky Way halo. In their model, the
agreement with the observed trend is a consequence of the fact that the stellar
halo originates from a few relatively massive satellites that were accreted
early on and were enriched in $\alpha$ elements by type II supernovae. The
model surviving satellites are accreted much later, have more extended star
formation histories and stellar population enriched to solar level by both type
II and type Ia supernovae. In the model by Font et al., the most metal-poor
stars are located in the inner 10 kpc, which is in contrast with the trends
found by Carollo et al. (2007).

The formation and structure of the stellar halo was studied, in the context of
a hybrid model of galaxy formation, in the work by De Lucia \& Helmi (2008)
mentioned above. The working hypothesis used in this study is that the stellar
halo built from the cores of the satellite galaxies that merged with the Milky
Way over its lifetime. In order to identify the stars that end up in the
stellar halo, the full merger tree of the model Milky Way galaxy was
constructed, and the galaxies that merge onto the main branch of the galaxy
identified. These galaxies were then traced back until they are for the last
time central galaxies, and a fixed fraction (ten per cent in the fiducial
model) of the most bound particles of their parent haloes were {\it tagged}
with the stellar metallicity of the galaxies residing at their centre.

\begin{figure}
\includegraphics[width=83mm,height=130mm]{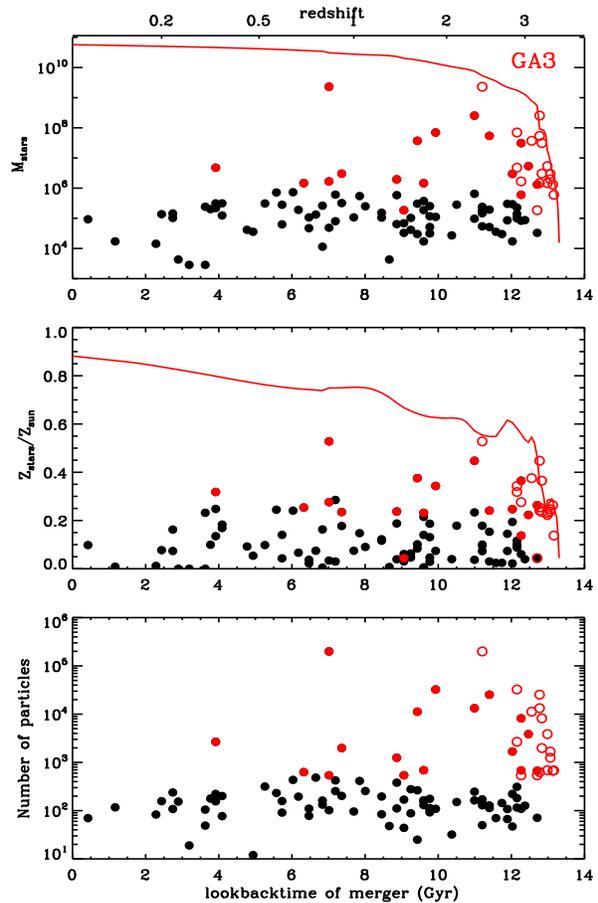}
\caption{From De Lucia \& Helmi (2008): stellar masses (top panel) and
  metallicities (middle panel) for all galaxies accreted onto the main branch
  for the simulation GA3, as a function of the lookback time of the galaxy's
  merger. The solid red lines in the top and middle panels show the evolution
  of the stellar mass and metallicity in the main progenitor of the Milky
  Way. The bottom panel shows the number of particles associated with the dark
  matter halo at the time of accretion, again as a function of the merging time
  of the galaxy that is located at its centre. Red symbols indicate objects
  associated with substructures with more than 500 particles. Red open symbols
  correspond to red filled circles but are plotted as a function of the time of
  accretion.}
\label{fig:accretedGA3}
\end{figure}

Fig.~\ref{fig:accretedGA3} shows the stellar masses (top panel) and
metallicities (middle panel) of all galaxies accreted onto the main branch for
the simulation GA3, as a function of the lookback time of the galaxy's
merger. Stellar masses and metallicities correspond to those at the time of
accretion.  The solid red lines in these panels show the evolution of the
stellar mass and of the stellar metallicity in the main progenitor of the Milky
Way galaxy (as in Fig.~\ref{fig:mgrowth}). The bottom panel of
Fig.~\ref{fig:accretedGA3} shows the number of particles associated with the
dark matter haloes before accretion, again as a function of the merging time of
the galaxies that reside at their centre. Red symbols indicate objects
belonging to haloes with more than 500 bound particles before accretion. For
these objects, both the time of accretion (open symbols) and the merger time
(solid symbols; this is given by the dynamical friction timescale) are
indicated in the figure. Most of the galaxies that merge onto the main branch
have stellar masses and metallicities that are much smaller than the current
mass and metallicity of the main progenitor, over most of the galaxy's
life-time. Most of the accreted galaxies lie in quite small haloes and only a
handful of them are attached to relatively massive systems.  These are the
galaxies that contribute most to the build-up of the stellar halo. The red
symbols in the bottom panel of Fig.~\ref{fig:accretedGA3} show that most haloes
with more than $\sim 500$ particles were all disrupted more than $\sim 6$~Gyr
ago. These haloes contain the few galaxies with stellar mass larger than
$10^6\,{\rm M}_{\odot}$ which merge onto the main branch over the galaxy's
life-time (top panel). The stellar metallicities are generally relatively low,
with a median value of $\sim 0.3\,{\rm Z}_{\sun}$, with larger values
associated with larger galaxies (see below).

The approach used by De Lucia \& Helmi is similar to that adopted by Bullock \&
Johnston (2005), although the two methods differ in a number of
details. Results by Font et al. (2007) are in good agreement with those
illustrated in Fig.~\ref{fig:accretedGA3}: they find that one or a few more
satellites in the range $10^8-10^{10}\,{\rm M}_{\odot}$ can make up 50-80 per
cent of the stellar halo, and that most of them are accreted early on (${\rm
  t}_{\rm accr} > 9$~Gyr). In agreement with findings by Bullock \& Johnston
(2005), the halo resulting from the simulations of De Lucia \& Helmi (2008)
exhibits a steeper profile and is more centrally concentrated than the dark
matter profile. In addition, they find that star particles with metallicity
larger than $0.4\,Z_\odot$ are more centrally concentrated than star particles
of lower abundances, in qualitative agreement with the measurements by Carollo
et al. (2007).

A more sophisticated tagging scheme has been recently used by Cooper et
al. (2010) who take advantage of the higher resolution simulations from the
Aquarius project. In their study, Cooper et al. assume that the energy
distribution of newly formed stars traces that of the dark matter. They then
order the particles by binding energy and select some fraction ($f_{\rm MB}$)
of these most bound particles to be tagged. $f_{\rm MB}$ is treated as a free
parameter, and is fixed by comparing model predictions with observational
measurements of the structure and kinematics of the Milky Way satellites. The
tagging scheme adopted by Cooper et al. is similar to that used in De Lucia \&
Helmi but, again, the details are different: De Lucia \& Helmi tag the most
bound ten per cent of particles in a satellite, but perform this tagging only
once (i.e. at the time when its parent halo becomes a subhalo of the main
halo). In contrast, Cooper et al. apply their tagging scheme each time a new
stellar population is formed and chose the subset of dark matter particles
according to the instantaneous dynamical state of the host halo.

\begin{figure}
\includegraphics[width=80mm,height=110mm]{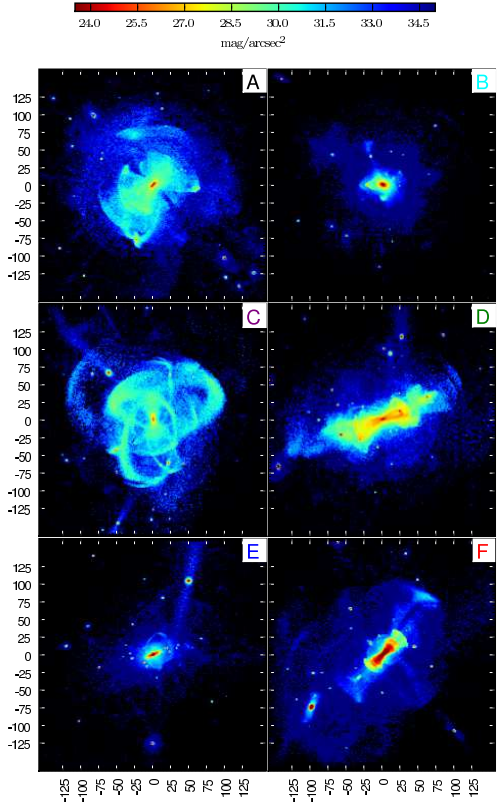}
\caption{From Cooper et al. (2010): $V$-band surface brightness of model
  stellar haloes (and surviving satellites), to a limiting depth of 35
  mag/arcsec$^2$. The axis scales are in kiloparsec.}
\label{fig:fig6cooper}
\end{figure}

Fig.~\ref{fig:fig6cooper} shows a $300\times300$~kpc projected surface
brightness map of the stellar halo at $z=0$, for each of the six Aquarius dark
matter haloes.  Substantial diversity among the six haloes is apparent. A few
haloes (e.g. Aq-B and Aq-E) are characterized by strong central concentrations,
while others show extended envelopes out to $75-100$~kpc. Each envelope is the
superposition of streams and shells that are phase-mixed to varying
degrees. Most haloes exhibit a strongly prolate distribution of stellar mass,
particularly in the inner regions. The brightest and most coherent structures
visible in Fig.~\ref{fig:fig6cooper} can be associated with the most recent
accretion events. In addition, Cooper et al. show that their model stellar
haloes span a wide range of accretion histories, ranging from a gradual
accretion of many progenitors (Aq-A, Aq-C and Aq-D) to one or two significant
accretions (Aq-B, Aq-E and Aq-F). All the most significant contributors to the
stellar haloes were accreted more than 8 Gyr ago, with the exception of
Aq-F. For this halo, more than 95 per cent of the halo was contributed by the
late merger of an object of stellar mass greater than the SMC infalling at
$z\sim 0.7$. Finally, only in two haloes (Aq-C, Aq-D), the stars stripped from
the surviving satellites represent the largest fraction (up to $\sim 70$ per
cent) of the stellar halo. In the other haloes, stars stripped from surviving
satellite represent less than $\sim 10$ per cent of the stellar halo.

\begin{figure}
\includegraphics[width=76mm,height=76mm]{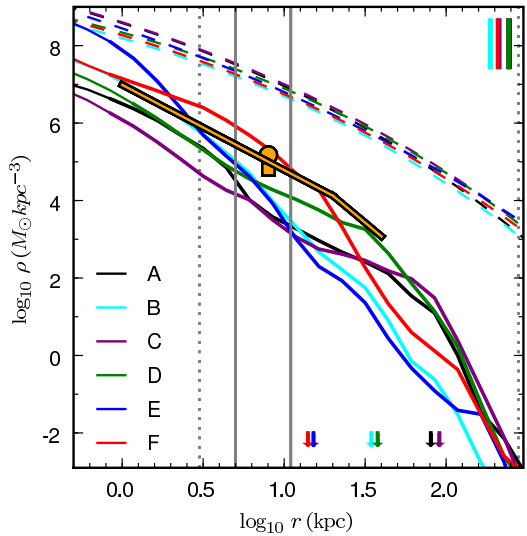}
\caption{From Cooper et al. (2010): Spherically averaged density profiles for
  the six Aquarius stellar haloes. Arrows mark the break radii of broken
  power-law fits to each profile. Dashed lines show Einasto profile fits to the
  corresponding dark matter haloes. Grey vertical lines demarcate what the
  authors define the `outer halo region' (dotted) and the `solar neighbourhood'
  (solid); coloured vertical bars indicate $r_{200}$ for the dark
  haloes. Overplotted are representative data for the Milky Way (orange):
  estimates of the halo density in the solar neighbourhood (symbols) from Gould
  et al. (1998, square) and Fuchs \& Jahrei\ss~(1998, circle), and the
  best-fitting broken power law of Bell et al. (2008, excluding the Sagittarius
  stream and Virgo overdensity).}
\label{fig:fig13cooper}
\end{figure}

Fig.~\ref{fig:fig13cooper} shows the spherically averaged density profiles of
halo stars for the six haloes analysed in Cooper et al. The density profiles
have been obtained by excluding material bound in surviving satellites, but
making no distinction between streams, tidal tails or other overdensities. Dark
matter haloes are well fitted by the Einasto profiles (shown as dashed lines in
the figure - see also Springel et al. 2008), while the stellar haloes exhibit a
notable degree of substructures which is due to the contribution of localized
and spatially coherent subcomponents within the haloes.  The \linebreak shapes
of the density profiles are broadly similar, showing a strong central
concentration and an outer decline which is steeper than that of the dark
matter halo. Cooper et al. note that the density profile of the many-progenitor
haloes (Aq-A, Aq-C and Aq-D) can be fit with broken power-law profiles, with
indices similar to the Milky Way ($n\sim 3$) interior to the break. In
contrast, two of the few-progenitor haloes (Aq-B and Aq-E) have steeper
profiles and show no obvious break, although their densities in the solar shell
are comparable to those of the many-progenitor haloes. Aq-F is dominated by a
single progenitor, and its debris retains a high degree of unmixed structure at
$z=0$.

In a follow-up study, Cooper et al. (2011) have compared predictions from their
models to the the two-point correlation function for Galactic halo stars (this
is based on a catalogue of blue horizontal branch stars identified in the Sloan
Digital Sky Survey). Considering only stars with $r>20$~kpc, five out of the
six Aquarius haloes show statistically significant correlation on scales
equivalent to $\sim 1$-$8$~kpc, and most of the models are consistent with the
observational data for the outer regions ($r>30$~kpc). For the inner regions of
the stellar halo, however, their simulations tend to be significantly more
clustered than the data. They argue that one possible explanation for this is
the existence of a smooth component that is not currently included in their
simulations, and that could originate from stars scattered from the disc or born
on eccentric orbits. It would be interesting to improve the available models
so as to include these physical processes.

\section{The Milky Way satellites}

\begin{figure}
\includegraphics[width=76mm,height=76mm]{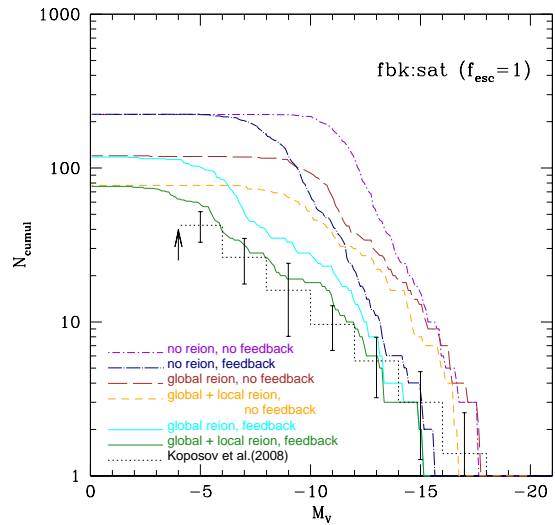}
\caption{From Font et al. (2011): The effect of different physical processes on
  the luminosity function of satellites in the Aq-A halo, in the fbk:sat model
  (see the original paper for details).}
\label{fig:fig7font}
\end{figure}

\begin{figure}
\includegraphics[width=76mm,height=76mm]{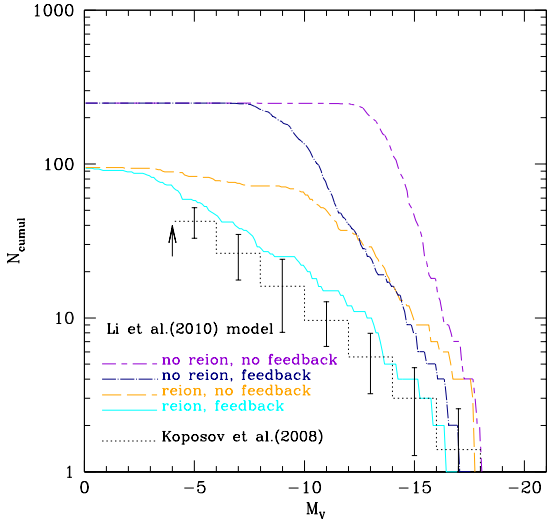}
\caption{From Font et al. (2011): Luminosity function for the Aq-A halo in the
  Li et al. (2010) model with global reionization and feedback switched on and
  off (see original papers for details).}
\label{fig:fig12font}
\end{figure}

As mentioned in Section~\ref{sec:intro}, it has long been realized that the
number and properties of the Milky Way satellites can be affected significantly
by baryonic physics, in particular by reionization and supernovae
feedback. Figs.~\ref{fig:fig7font} and \ref{fig:fig12font} show the effect of
these physical processes on the luminosity function of Milky-Way satellites,
from two independently developed semi-analytic models both applied to the
second highest level of resolution of the Aq-A halo.

Fig.~\ref{fig:fig7font} shows results from the model described in Font et
al. (2011) who implemented a detailed treatment of reionization in the Durham
semi-analytic model of galaxy formation, {\small GALFORM}. In this model, the
UV flux produced by the galaxy population is calculated in a self-consistent
way, and the contribution from quasars is taken from the observationally
inferred spectrum of Haardt \& Madau (1996). The model also allows for an
additional local UV flux generated by the progenitors of the Milky Way
galaxy. This results in an earlier effective redshift of reionization which
suppresses the star formation in satellite galaxies, particularly in very low
mass galaxies.  In order to achieve a good agreement with observational
measurements, the model requires, however, an escape fraction of UV photons of
about 100 per cent. The figure shows that, in this model, the number of faint
satellites is reduced by a combination of supernovae feedback and global+local
photoheating. Supernovae feedback is effective in suppressing the more massive
systems, while photoheating plays an important role in reducing the number of
ultra-faint dwarfs. Local photoheating appears to be a crucial ingredient.

Fig.~\ref{fig:fig12font} shows the corresponding results from the model
presented in Li et al. (2010), applied to the same dark matter simulation. This
model is based on that described in De Lucia \& Helmi (2008) but has been
updated to provide a better match to the observed properties of Milky Way
satellites. It differs from the model by Font et al. in a number of details. In
particular: a simple reionization model, that does not account for local
photoionization is adopted, based on the Gnedin (2000) formalism. Cooling via
molecular hydrogen is not included, under the assumption that ${\rm H}_2$ is
efficiently photodissociated. All other physical processes considered are
modelled using different prescriptions (see original papers for
details). Fig.~\ref{fig:fig12font} shows that the relative importance of
reionization and feedback is similar to that found in the Font et al. model:
reionization alone is not able to bring the predicted number of luminous
satellites in agreement with that observed, and is effective at the faint end
of the luminosity function ($M_{V} \gtrsim -10$). Feedback from supernovae is
instead important for more luminous satellites. 

Similar conclusions have been reached also by Macci\`o et al. (2010) who
compare results from three independently developed semi-analytic models of
galaxy formation. The models are applied to high-resolution $N$-body
simulations but do not use (as in the studies discussed above) the subhalo
information to follow the evolution of the satellite population. Several other
studies of the Milky Way satellites in a $\Lambda$CDM framework have been
carried out recently. A brief overview of the most recent ones can be found in
Appendix B of Font et al. (2011). Both Li et al. (2010) and Font et al. (2011)
have compared predictions of their models with other physical properties of the
Milky Way satellites. In particular, they have shown that both models are able
to provide a relatively good agreement with the observed cumulative radial
distribution of satellites and the observed metallicity-luminosity
relation. Again, one important limitation of these studies is that they both
assume an instantaneous recycling approximation for chemical enrichment.

\begin{figure}
\includegraphics[width=80mm,height=50mm]{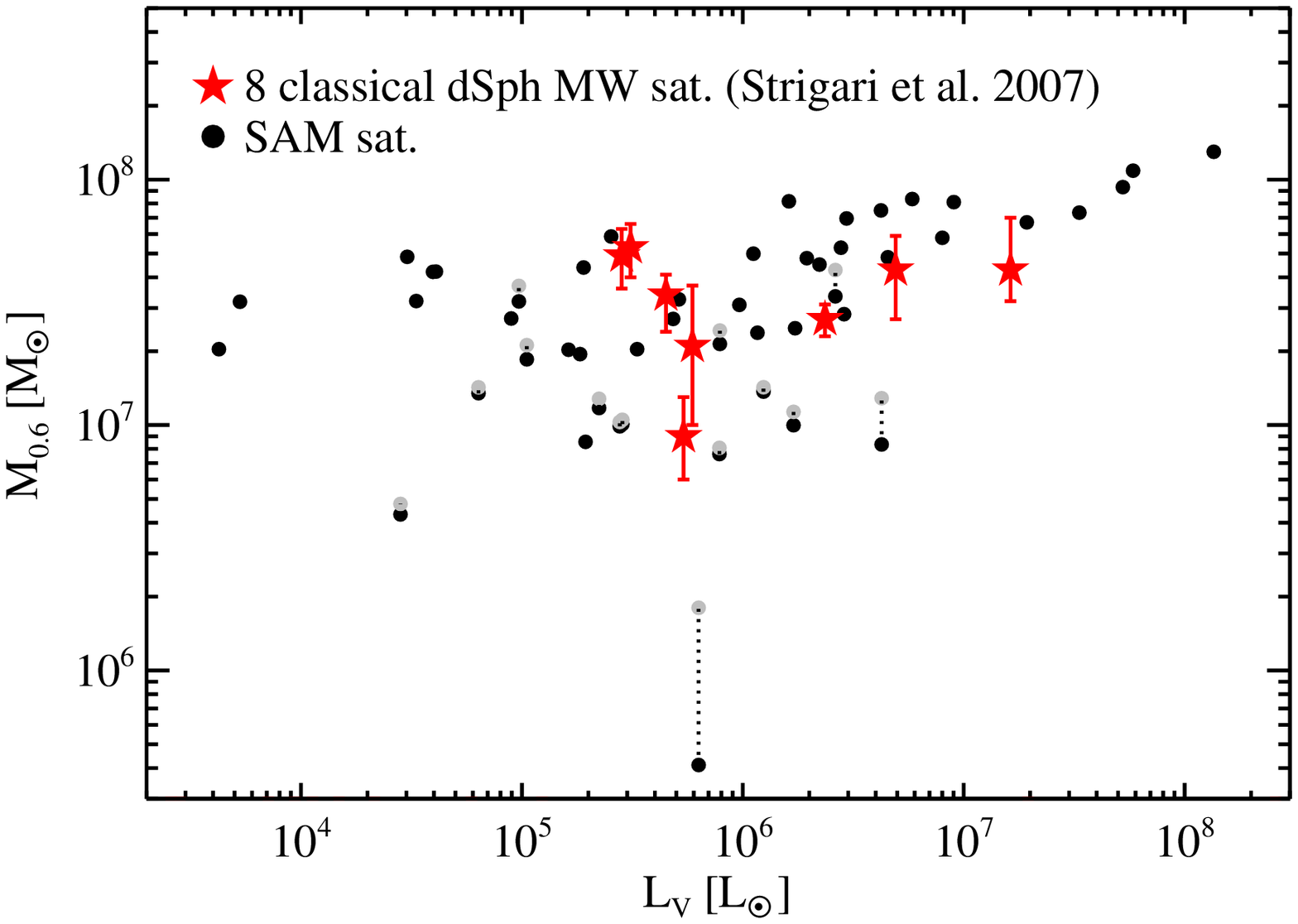}
\includegraphics[width=80mm,height=50mm]{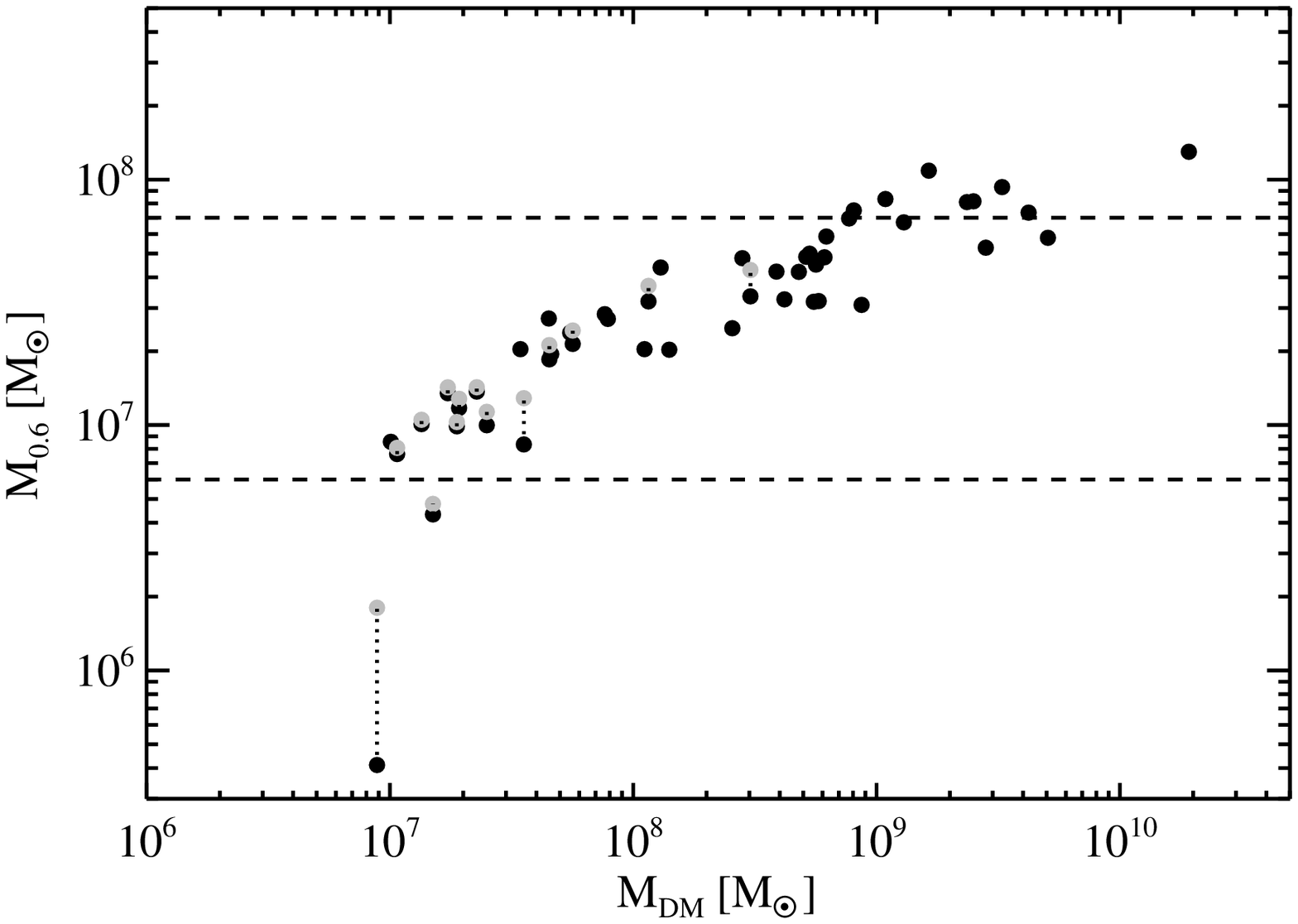}
\caption{From Li et al. (2009): The top panel shows the mass within 600 pc as a
  function of the $V$-band luminosity for model satellites, and for the eight
  brightest dwarf spheroidal galaxies of the Milky Way (asterisks). Black and
  grey circles corresponds to two alternative measurements of ${\rm M}_{0.6}$
  (see original paper for details). The bottom panel shows the mass within 600
  pc as a function of the total mass of the dark matter substructure. The two
  dashed lines correspond to the upper and lower limits for the observational
  estimates by Strigari et al. 2008.}
\label{fig:figli}
\end{figure}

The high resolution of the simulations used in the studies mentioned above
allows the satellites dark matter mass enclosed within a (small) given radius
to be measured. Fig.~\ref{fig:figli} shows how the mass within 600 pc
($M_{0.6}$) varies as a function of the $V$-band luminosity of model satellites
in the top panel, and as a function of the total mass of the dark matter
substructure in the bottom panel. Li et al. 2009 did not attempt to measure the
mass within 300 pc (as done in more recent observational studies) as this would
go beyond the resolution limit of the simulation used in their study (the
highest resolution simulation of the GA series). The figure shows that the
correlation between $M_{0.6}$ and $V$-band luminosity is somewhat stronger than
in the real data, but consistent with them. The bottom panel of
Fig.~\ref{fig:figli} shows that the present day total mass of substructures
hosting luminous satellites varies in the range $\sim 10^7$-$10^{10}\,{\rm
  M}_{\odot}$. The scatter reflects the lack of a tight concentration-virial
mass relation, the different accretion times, and the different amounts of
tidal stripping suffered by the parent substructures once they have fallen on
to the Milky Way halo. Therefore, in the model by Li et al., a common mass
scale within 600 pc for luminous satellites results from the strong suppression
of accretion and cooling of gas in low mass haloes after reionization, as well
as from the atomic hydrogen cooling threshold at $T_{\rm vir} = 10^4\,{\rm
  K}$. Similar conclusions have been obtained in more recent studies that have
taken advantage of higher resolution simulations to measure the mass within a
radius of 300 pc. Interestingly, all these studies have found a weak dependence
of $M_{300}$ on luminosity (e.g. Macci\`o et al. 2009, Mu\~noz et al. 2009,
Busha et al. 2010, Font et al. 2011). This represents then a robust and
testable prediction that requires improved measurements of $M_{300}$.

\section{Things we have learned}
The work discussed above has demonstrated that, within the current standard
paradigm for structure formation, it is possible to reproduce the basic
properties of our Milky Way and of its satellite systems using {\it plausible}
prescriptions to model the relevant physical processes at play.

In agreement with earlier studies, recent work has confirmed that combining a
sufficiently high redshift of reionization with a relatively strong feedback
from supernovae, it is possible to bring the predicted number of luminous
satellites in agreement with the most recent observational results. The same
models provide a relatively good agreement with some basic physical properties
measured for the Milky Way satellites. In addition, the models also provide an
explanation for the apparent common mass scale of dwarf galaxies, which results
from a weak dependence of $M_{300}$ on the virial mass of the substructures
hosting luminous galaxies. In fact, models predict a weak increase of $M_{300}$
for increasing luminosity which will be testable once more accurate
measurements are available.

Model results depend strongly on the particular implementation of supernovae
feedback, which is effective in \linebreak suppressing the formation of
relatively luminous satellites. A scheme which is more efficient in galaxies
residing in lower mass haloes provides, however, a better agreement with the
properties of the ultra-faint satellites. In the framework of these models, the
brightest satellites are associated with the most massive subhaloes and are
accreted relatively recently ($z<1$). They have extended star formation
histories and only a small fraction of their stars are made by the end of the
reionization.  On the other hand, fainter satellites tend to be accreted early,
are dominated by old stellar populations, and a number of them formed most of
their stars before reionization was complete.

The stellar halo is built predominantly by accretion of satellites at early
times. A range of accretion histories is possible, from smooth growth of the
stellar halo to growth in one or two discrete events. The most significant
accretion events are those that occurred more recently, and they dominate the
stellar halo at large radii. The halo has a complex structure, with well-mixed
components, tidal streams, and shells, that is not well described by smooth
models.

In these models, the stars in the halo do not exhibit any significant
metallicity gradient, but higher metallicity stars are more centrally
concentrated than stars with lower abundances. \linebreak This is due to the
fact that the most massive satellites contributing to the stellar halo are also
the most metal rich, and dynamical friction drags them closer to the inner
regions of the host halo. Finally, in the context of these models, the observed
abundance pattern of the stellar halo and of the surviving dwarf spheroidals
can be reproduced.

\section{Open issues}

However, we are not left without problems, and open questions. 

Most of the recently published hybrid models (including all those discussed in
the previous sections) adopt an instantaneous recycling approximation which is
inappropriate for the iron-peak elements, mainly produced by supernovae Type
Ia. This is a crucial missing ingredient in order to understand if the models
discussed above are really successful (see e.g. discussion related to
Fig.~\ref{fig:metdistr}). Work is ongoing in order to improve the available
models by taking into account time-dependent yields and following explicitly
the evolution of different elements. This will allow a more direct and thorough
comparison with available observational data to be carried out, thereby
providing even stronger constraints on available models.

At least for some of the models discussed above, the predicted relation between
stellar mass and dark matter mass (at the time of infall for satellite
galaxies) is offset with respect to that obtained using abundance matching
techniques (Guo et al. 2010). This goes in the sense that at a given
luminosity, model galaxies reside in smaller dark matter haloes. The difference
is largest at small masses where observational constraints are poor, but it
persists also at scales of the Milky Way galaxy (Starkenburg et al., in
preparation). Guo et al. (2011) have recently presented an extension of
previous models that shows a good agreement with the relation derived from
abundance matching. In order to achieve this match, the model adopts a
supernovae feedback ejection efficiency that depends strongly on the velocity
of the parent dark matter halo. The predicted luminosity function of Milky Way
satellites is marginally inconsistent with the observed number of faint
satellites, although still plausible given the observational uncertainties. It
will be interesting to check if this model also reproduces the detailed
physical and chemical properties of the Milky Way satellite galaxies.

\begin{figure}
\includegraphics[width=83mm,height=76mm]{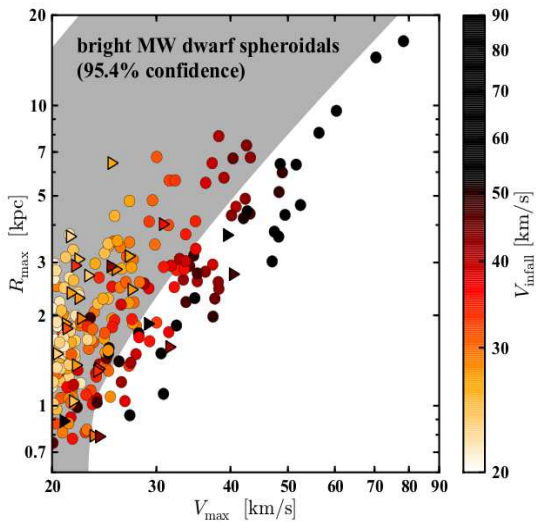}
\caption{From Boylan-Kolchin et al. (2011a): subhaloes from all six Aquarius
  simulations (circles) and from the Via Lactea II simulation (triangles),
  colour-coded according to $V_{\rm infall}$. The grey-shaded region shows the
  2$\sigma$ confidence interval for possible hosts of the bright Milky Way
  dwarf spheroidals (see original paper for details).}
\label{fig:fig2bk}
\end{figure}

Recently, Boylan-Kolchin et al. (2011a) have shown that the state-of-the-art
collisionless $N$-body simulations predict a population of massive dark matter
substructures that are too concentrated (have too high a $V_{\rm max}$ for a
given $r_{\rm vmax}$) to be able to host the brightest satellites of the Milky
Way. The problem is illustrated in Fig.~\ref{fig:fig2bk} that shows the
location in the $V_{\rm max}$ versus $r_{\rm vmax}$ plane for subhaloes
extracted from the six Aquarius haloes and from the via Lactea II simulation,
and the constraints from the Milky Way dwarf spheroidal galaxies. This poses a
problems for abundance matching methods that, while reproducing the luminosity
function of the Milky Way satellites, they do so by assigning the brightest
satellites to haloes that are denser than observed.  In a follow-up paper,
Boylan-Kolchin et al. (2011b) have explored possible solutions to these
problems, in the context of $\Lambda$CDM and argued that all of them appear to
be unlikely. In particular, they have analysed the following options: (1) an
atypical Milky Way halo that is either deficient in massive subhaloes or
populated by atypical substructures; (2) a stochastic galaxy formation at small
scales that washes out the correlation between halo mass and luminosity; (3)
strong baryonic feedback that reduces the central density of massive subhaloes
by large amounts. If all of the `astrophysical solutions' are rejected, we are
left with the more fundamental problem of the nature of dark matter. For
example, in a very recent work, Lovell et al. (2011) argue that no such problem
exists if haloes are made of warm dark matter.

Finally, it remains to be verified that a good agreement with the number and
also the physical properties of the \linebreak dwarf satellites of the Milky
Way can be retained while simultaneously matching the properties of the global
galaxy population at different cosmic epochs.

\acknowledgements 

I thank the organizers of the AG2011 meeting for the invitation, and for the
friendly and stimulating atmosphere. I acknowledge financial support from
the European Research Council under the European Community's Seventh Framework
Programme (FP7/2007-2013)/ERC grant agreement \linebreak n.~202781.


\end{document}